\documentclass[11pt, executivepaper]{article}
\usepackage[utf8]{inputenc}
\usepackage[T1]{fontenc}
\usepackage{natbib}
\usepackage{amsmath}
\usepackage{mathtools}
\usepackage{xcolor}
\usepackage{amsfonts}
\usepackage{graphicx}
\usepackage{enumitem}
\usepackage{geometry}
\usepackage{hyperref}
\hypersetup{colorlinks= true, allcolors=blue}
\setcitestyle{aysep={}}
\begin{document}

\title{\textbf{Some remarks on the mentalistic reformulation of the measurement problem. A reply to S. Gao}}

\author{Andrea Oldofredi\thanks{Contact Information: Universit\'e de Lausanne, Section de Philosophie, 1015 Lausanne, Switzerland. E-mail: Andrea.Oldofredi@unil.ch}}

\maketitle

\begin{abstract}
\cite{Gao:2017aa} presents a new mentalistic reformulation of the well-known measurement problem affecting the standard formulation of quantum mechanics. According to this author, it is essentially a determinate-experience problem, namely a problem about the compatibility between the linearity of the Schr\"odinger's equation, the fundamental law of quantum theory, and definite experiences perceived by \emph{conscious observers}. In this essay I aim to clarify (i) that the well-known measurement problem is a mathematical consequence of quantum theory's formalism, and that (ii) its mentalistic variant does not grasp the relevant causes which are responsible for this puzzling issue. The first part of this paper will be concluded claiming that the ``physical'' formulation of the measurement problem cannot be reduced to its mentalistic version. In the second part of this work it will be shown that, contrary to the case of quantum mechanics, Bohmian mechanics and GRW theories provide clear explanations of the physical processes responsible for the definite localization of macroscopic objects and, consequently, for well-defined perceptions of measurement outcomes by conscious observers. More precisely, the macro-objectification of states of experimental devices is obtained exclusively in virtue of their clear ontologies and dynamical laws \emph{without} any intervention of human observers. Hence, it will be argued that in these theoretical frameworks the measurement problem and the determinate-experience problem are logically distinct issues. 
\vspace{4mm}

\vspace{5mm} 
\begin{center}
\emph{Accepted for publication in Synthese}
\end{center}
\end{abstract}
\clearpage

\tableofcontents
\vspace{10mm}

\section{Introduction: A ``mentalistic'' version of the Measurement Problem}

\cite{Gao:2017aa} presents a new mentalistic reformulation of the well-known Measurement Problem (MP) affecting the standard formulation of Quantum Mechanics (QM). According to this author, the MP is essentially a determinate-experience problem, namely a problem about the compatibility between the linearity of the Schr\"odinger's Equation (SE), the fundamental law of quantum theory, and the definite experiences perceived by \emph{conscious observers}.
More precisely, Gao states that the physical formulation of the issue contained in \cite{Maudlin:1995aa}, considered a precise description of the problem in the philosophical literature, does not capture the substantial core of the MP, for it literally ignores the existing psycho-physical connections between definite perceptions of measurements outcome by conscious observers on the one hand, and the linear dynamics of the SE on the other. Therefore, the MP should be properly characterized as a determinate-experience problem:
\begin{quote}
[t]he problem is not only to explain how the linear dynamics can be compatible with the appearance of definite measurement results obtained by physical devices, but also, and more importantly, to explain how the linear dynamics can be compatible with the existence of definite experiences of conscious observers (\cite{Gao:2017aa}, p. 4).
\end{quote}
\noindent It is worth noting that in his essay Gao makes two strong statements (among others):\footnote{I assume that the reader is familiar with the content of Gao's and Maudlin's papers.} he writes not only that a mentalistic version of the MP is more ``appropriate'' to better understand the real problem of measurement in the context of quantum theory, but also that Maudlin's formulation is \emph{reduced} to his mentalistic variant (``Maudlin's formulation will reduce to the new formulation'', \cite{Gao:2017aa}, p. 5). Hence, in order to evaluate these claims, let us introduce the mentalistic measurement problem, which arises from the incompatibility among the following statements:
\begin{enumerate}
   \item The mental state of an observer supervenes on her wave function;
   \item The wave function evolves according to the SE;
   \item A measurement yields a single mental state representing a single outcome.
\end{enumerate}

The proof of this inconsistency is given in \cite{Gao:2017aa}, p. 4 and follows accurately the logical steps of \cite{Maudlin:1995aa}, p. 7-8. 

The different physical solutions to the MP, Bohmian Mechanics (BM), Many Worlds Interpretation (MWI) and the spontaneous collapse theories (here we will consider the original Ghirardi-Rimini-Weber (GRW) theory and its extensions GRWm and GRWf, which postulate a matter density field and a flash ontology respectively) represent in Gao's account diverse forms of psychophysical connections. Against this background, the author analyzes each of them in order to establish (i) whether and how these solutions satisfy the principle of psychophysical supervenience, and (ii) how observers' mental states supervene on their wave function. It follows, then, that a satisfactory solution of the mentalistic MP will have to explain how a conscious observer will perceive a definite experience of a given measurement outcome. Alternatively stated, Gao requires from a satisfactory solution to the MP that the physical state obtained as a measurement result must be the physical state on which the mental state of a conscious observer supervenes. Analyzing these different sorts of psychophysical connection and their implications Gao argues that 
\begin{quote}
the forms of psycho-physical connection required by Everett's and Bohm's theories have potential problems, while an analysis of how the mental state of an observer supervenes of her wave function may help to solve the structured tails problem of collapse theories. This seems to suggest that collapse theories may be in the right direction to solve the measurement problem (\cite{Gao:2017aa}, p. 11).
\end{quote}

The aim of this essay is twofold: on the one hand, I will criticize Gao's starting assumption according to which the MP is inherently a determinate-experience problem, and thus I will object that his mentalistic variant is \emph{more} appropriate with respect to the physical formulation of the issue, rejecting also the reduction of the MP to his new formulation (Section \ref{MP}). On the other, it will be shown that, contrary to the case of standard quantum mechanics, BM and GRW theories provide clear explanations of the physical processes responsible for the definite localization of macroscopic objects and, consequently, for well-defined perceptions of measurement outcomes by conscious observers (Sections \ref{BM} and \ref{GRW}).\footnote{In this paper I will concentrate only on Bohmian mechanics and GRW theories since they have a clear primitive ontology of matter, contrary to MWI.} More specifically, I will argue that looking carefully at their physical content, measurement results - e.g. macroscopic states of classical experimental devices - are obtained unambiguously from the microscopic dynamics \emph{without} any intervention of human observers. Furthermore, these theories guarantee, in virtue of their Primitive Ontology (PO) and dynamical laws, that observers' mental states will supervene on well localized physical states representing measurement outcomes, in agreement with Gao's requirements stated above. Consequently, observers' conscious perceptions of macroscopic states are always well-defined in BM and GRW.

Finally, I will argue that in BM and GRW theories the physical formulation of the MP is not reduced to this new mentalistic version. The latter, ultimately, is not relevant in order to evaluate and understand how these theories effectively solve the MP. The last section concludes the paper.

\section{The Problem of Measurement in Quantum Mechanics}
\label{MP}

Generally speaking quantum mechanics may be defined as a theoretical framework providing an algorithm able to describe the objects composing our world at the microphysical regime, i.e. atoms, molecules, subatomic particles such nucleons or electrons, and their motion in space and time. 

According to the von Neumann-Dirac formulation of the theory, a quantum system is described by a state vector $|\psi\rangle$ which is an element of a complex vector space, the Hilbert space $\mathcal{H}$, providing a \emph{complete} specification of its properties. Equivalently, considering Schr\"odinger's wave mechanics, the physical state of a quantum object is described by a wave function $\psi$, a complex valued function defined in $3N$-dimensional configuration space. The wave function of a given system contains all information about its state, providing its \emph{complete} description. In QM only \emph{operationally accessible} (i.e. measurable) properties are considered magnitudes of quantum systems; these are represented by Hermitian operators $\mathcal{A}$. The possible values of a given measurable quantity $A$ are real numbers called \emph{eigenvalues} of the associated operator $\mathcal{A}$, whereas possible states in which a system may be found after a measurement of $A$ are represented by the eigenstates/eigenvectors of $\mathcal{A}$. This is the well-known eigenvalue-eigenstate link, a central tenet of quantum mechanics. Dirac's wrote illuminating words on this point:
\begin{quote}
The expression that an observable `has a particular value' for a particular state is permissible in quantum mechanics in the special case when a measurement of the observable is certain to lead to the particular value, so that the state is in an eigenstate of the observable $[\dots]$. In the general case we cannot speak of an observable having a value for a particular state, but we can speak of its having an average value for the state. We can go further and speak of the probability of its having any specified value for the state, meaning the probability of this specified value being obtained when one makes a measurement of the observable (\cite{Dirac:1947aa}, p. 253).
\end{quote}
This quote eloquently characterizes the idea that quantum systems in isolation, i.e. not subject to experimental observations, do \emph{not} instantiate properties with definite values. Consequently, it follows from the formalism of the theory that quantum mechanical objects possess \emph{indefinite} properties prior measurements, a crucial ontological difference with respect to their classical counterparts. 

The last axiom of QM needed in our discussion concern the dynamics of physical systems. The evolution in space and time of quantum systems is governed by the already mentioned Schr\"odinger's equation, a deterministic linear partial differential equation:
\begin{align}
\label{SE}
i\hbar\frac{\partial}{\partial{t}}|\psi\rangle =\hat{H}|\psi\rangle,
\end{align} 
\noindent{where} $i$ is the imaginary unit, $\hbar$ the reduced Planck constant and $\hat{H}$ is the Hamiltonian operator representing the total energy of the system. The Hamiltonian operator is particularly important being the generator of the time evolution of quantum systems. 

Among the properties of SE, linearity plays a major role for the definition of the MP: if two state vectors $|\psi_1\rangle,|\psi_2\rangle$ (or wave functions $\psi_1, \psi_2$) are both possible solutions of the same SE, then their linear combination (\emph{superposition}) $|\psi\rangle_s=\alpha|\psi_{1}\rangle+\beta|\psi_{2}\rangle$ ($\psi_s=\alpha\psi_1+\beta\psi_2$) is still a solution of the same SE.\footnote{Here $|\alpha|^2,|\beta|^2$, with $\alpha,\beta\in\mathbb{C}$, represent the probabilities to find the system in $|\psi_1\rangle,|\psi_2\rangle$ respectively. The normalization $|\alpha|^2+|\beta|^2=1$ means that with certainty we will find the system in one of the possible eigenstates.} Thus, the new superposed state $|\psi\rangle_s$ is also a consistent representation of the system. 

Now suppose to measure the $z$-spin of a quantum particle. In this particular case there are only two admissible eigenstates ``$z$-spin-up'' and ``$z$-spin-down''; from linearity, nonetheless, it follows that also the superposition ``$z$-spin-up \emph{and} $z$-spin-down'' is a consistent description of the state in which the particle may be (i.e. it is completely admitted by the theory's formalism as a proper description of a physical state of a certain system). This entails that prior to a spin measurement the particle has \emph{indefinite} spin, being neither in the $z$-spin-up state \emph{nor} in the $z$-spin-down state; formally, this situation can be translated into the familiar equation:
\begin{align}
\label{spin}
|\psi_z\rangle=\frac{1}{\sqrt{2}}(|\uparrow_z\rangle+|\downarrow_z\rangle).
\end{align}

As we will see, assuming (i) that state vectors/wave functions provide a complete description of quantum systems and (ii) that their motion is uniquely and completely described by the SE, the superposed states would have been amplified to the \emph{macroscopic scale}. Since also the behaviour of the macro-apparatus, the experimental device, must obey the fundamental laws of QM being constituted by a huge number of quantum particles, it follows that macroscopic objects can be in superposition, contradicting empirical evidence, i.e. uniqueness and definiteness of measurement outcomes. This is in essence of the famous measurement problem of quantum theory. Schematically, it may be stated as follows:
\begin{enumerate}
   \item[\emph{P. 1}] State vectors provide a complete description of quantum systems;
   \item[\emph{P. 2}]State vectors evolve according to the linear dynamical equation \eqref{SE}; 
   \item[\emph{P. 3}]Measurements have a unique determinate outcome.
\end{enumerate}

Any pair of these propositions is consistent and entails the falsity of the third one, but their conjunction generates inconsistencies with experimental evidence. For a simple proof of this statement the reader should refer to \cite{Maudlin:1995aa}, p. 7-8. It is worth noting that observers do not play any role in the definition of the measurement problem. In the remainder of this section it will be made clear where they enter into the scene.

In order to show explicitly how quantum superpositions amplify to the macroscopic regime, let's consider again equation \eqref{spin}. Suppose that prior to the measurement of the particle's spin the in $z$-direction we have the following situation:
\begin{align*}
\sum_{i}c_i|\psi_i\rangle\otimes|A_0\rangle,
\end{align*}

\noindent where $\sum_{i}c_i|\psi_i\rangle$ refers to the two possible eigenstates in the r.h.s. of \eqref{spin}, and $|A_0\rangle$ to the experimental apparatus in its ready state, i.e. a macroscopic pointer pointing in a neutral direction before the measurement. Due to the linearity of \eqref{SE}, one obtains the following result as a mathematical consequence:

\begin{align}
\label{superpos}
\sum_{i}c_i|\psi_i\rangle\otimes|A_0\rangle\Longrightarrow\sum_{i}c_i|\psi_i\rangle\otimes|A_i\rangle
\end{align}

\noindent where $\sum_{i}c_i|\psi_i\rangle\otimes|A_i\rangle$ represent the entangled state between the quantum system and the experimental device. It is crucial to underline that equation \eqref{superpos} does neither describe a macroscopic apparatus pointing to a definite direction $i$, nor a quantum object with a definite state/value for the property of spin in the $z$-direction. Only when the experimenter decides to perform a measurement of the particle's spin the superposition present in \eqref{superpos} will collapse.

However, since in experimental observations such superpositions are never revealed, the von Neumann-Dirac formulation of quantum mechanics introduces a second stochastic dynamical law: the postulate of the \emph{projection} of the state vector, also known as the wave function collapse. Essentially, given the structure of the MP presented above, one may state that the standard solution offered by this approach to QM is to reject \emph{P. 2}, according to which the dynamical behaviour of quantum systems is \emph{completely} described by the SE. In our example, the act of observation of the particle's $z$-spin - i.e. the interaction between the quantum particle and the experimental apparatus - causes a suppression of the Schr\"odinger's evolution with the consequent stochastic jump of the system in one of the two possible spin eigenstates. These stochastic ``jumps'' make QM inherently probabilistic. Dirac viewed them as ``unavoidable disturbance'' of quantum systems in measurement situations:
\begin{quote}
When we measure a real dynamical variable $\xi$, belonging to the eigenvalue $\xi'$, the disturbance involved in the act of measurement causes a jump in the state of the dynamical system. From physical continuity, if we make a second measurement of the same dynamical variable immediately after the first, the result of the second measurement must be the same as the first. Thus after the first measurement has been made, there is no indeterminacy in the result of the second. Hence after the first measurement is made, the system is in an eigenstate of the dynamical variable $\xi$, [$\dots$]. In this way, we see that a measurement always causes the system to jump into an eigenstate of the dynamical variable that is being measured, the eigenvalue this eigenstate belongs to being equal to the result of the first measurement (\cite{Dirac:1947aa}, p. 36).
\end{quote}

Clearly, the projection postulate is essential to reconcile the postulates of quantum theory with experimental evidence and practice. In fact, stating that quantum systems are subjected to stochastic jumps when they interact with measurement devices, causing the instantaneously suppression of the unitary evolution, implies that these interactions have the macroscopic effect to correlate the state of the experimental apparatus with the eigenstate (and relative eigenvalue) in which the quantum system will be found, so that one can actually observe definite measurement results. This is the content and enormous utility of the projection postulate. 

Without this stochastic dynamical law QM would be inadequate to describe measurement outcomes, being able only to describe the evolution of physical systems in isolation (a very limited success for a fundamental physical theory). Nonetheless, although the projection postulate makes the quantum formalism consistent with experimental evidence, it does not provide a \emph{satisfactory} solution to the MP. In the first place, it notoriously implies an indispensable, arbitrary and not precisely defined division between the micro- and macroscopic regimes. Secondly, the notion of measurement, albeit pivotal within the axioms of QM and taken as an unexplained primitive concept, is neither mathematically nor physically well-defined, i.e. there are no variables in the equations of the theory referring the notions of measurement or observer (for a detailed analysis of these issues the reader may refer to \cite{Bell:2004aa}). In the third place, the physical processes which causes the collapse of the wave functions are not described, with the consequence that interactions occurring in measurements processes do not receive a physically meaningful explanation. More precisely, \emph{the physical processes} responsible for the suppression of the Schr\"odinger evolution are left unexplained: the stochasticity of quantum jumps, although efficient, seems to be a pragmatical rule introduced \emph{ad hoc} in order to tame inconsistencies with respect to observed experimental facts. In addition, there is no explanation of what distinguishes a measurement interaction from a non-measurement interaction:
\begin{quote}
[w]hat the traditional theory did \emph{not} do is state, in clear physical terms, the conditions under which the non-linear evolution takes place. There were, of course, theorems that if one puts in collapses \emph{somewhere} between the microscopic and the macroscopic, then, for all practical purposes, it doesn't much matter \emph{where} they are put in. But if the linear evolution which governs the development of the fundamental object in one's physical theory occasionally \emph{breaks down} or \emph{suspends itself} in favor of a radically different evolution, then it is a physical question of the first order exactly under what circumstances, and in what way, the breakdown occurs. The traditional theory papered over this defect by describing the collapses in terms of imprecise notions such as ``observation'' or ``measurement'' (\cite{Maudlin:1995aa}, p. 9).
\end{quote}

\noindent Finally, the nature of quantum objects is ontologically obscure. From QM it follows that generally prior to any measurement, the operator associated to the observable property under consideration does not have any fixed value. Thus, one concludes that quantum systems have indefinite properties before the performance of a measurement of a given observable. Alternatively stated, we might better say that the measurement in quantum mechanics does not reveal existing properties of quantum systems. Thus, quantum systems are objects whose properties depend \emph{contextually} on the different experimental situations. 
\vspace{2mm}

\noindent From our discussion of the measurement problem it is possible to draw some conclusions:
\begin{itemize}
   \item If one accepts the validity of the quantum mechanical postulates and their standard interpretation, i.e. the formal structure of the theory, the MP is just a direct \emph{mathematical} consequence deriving from the amplification of microscopic superpositions to the macroscopic scale. This, in turn, translates into a \emph{physical} problem since these macroscopic superpositions are never observed, making the theory inadequate to describe empirical evidence. It is worth noting that, in order to define the problem of measurement, there is no necessity to introduce a conscious observer. Hence, it seems that the physical formulation of the MP cannot be simply \emph{reduced} to its mentalistic variant. The latter does not seem to be the correct version of the problem, since the MP does \emph{not} regard conscious perceptions of human beings but rather the microscopic physical processes which generate definite macroscopic states of the experimental devices. 
    \item The collapse postulate affirms that the interactions occurring performing measurements suppress the Schr\"odinger evolution and cause the reduction of the wave packet in one of the possible eigenstates of the measured observable. The stochastic jump of the wave function, thus, generates a \emph{definite} macroscopic outcome. Hence, it is certain that the physical state on which the mental state of a conscious observer supervenes is macroscopically well-defined: it is the very act of observation which causes the definiteness of the macroscopic apparatus. Clearly, then, an observer will have a definite experience of the measurement outcome. More precisely, if one interprets literally the projection postulate, then one is forced to accept that the act of observation by an observer instantaneously causes the suppression of the Schr\"odinger evolution, i.e. the act of observation is \emph{simultaneous} to the macro-objectification of a definite state of the experimental device. Thus, the brain of a human observer does not have the physical time to observe a superposition of macroscopic states. Hence, it follows that an observer observes a definite macroscopic state also in the context of standard quantum mechanics. This fact, in turn, implies that the macroscopic physical state obtained after a measurement is exactly the state of which an observer has a direct perception. By definition of the projection postulate, then, both the macroscopic state and the observer's conscious perception will be well-defined. The difficulties related to this postulate are expressed above, but they do not concern the mental states of conscious obersers. Trivially, without projection postulate one has to face also a ``mentalistic MP'': if we accept \emph{P. 1, P. 2} and \emph{P. 3}, then we have to explain also how a human observer will have a definite perception of a well-defined measurement outcome. 
     \item Although historically the consciousness of a human observer has been introduced by von Neumann and Wigner in order to break the macroscopic superposition present in the r.h.s. of \eqref{superpos}, one has to keep in firmly mind that a human observer is only \emph{one type of observer}, as Bell and Feynman among others pointed out. In fact, there is no clear definition of what an observer may be: it could be any thing that performs measurements, e.g. a Geiger counter, not necessarily a rational agent with mental states.
\end{itemize}

In conclusion, it seems that the MP is strictly related to the macro-objectification of well-defined physical states from a microscopic not clear dynamics, rather than related to conscious human experiences. Therefore, Gao's analysis does not provide a better, or more exact exposition of the issue, since it lacks a clear discussion of the formal and physical reasons which cause the MP in the first place.

In the next sections we will analyze how Bohmian mechanics and the spontaneous collapse theories solve this foundational issue. In these contexts observers (human or not) do not play \emph{any} role since the MP is solved in virtue of their clear ontologies and dynamical laws. Furthermore, we will see how in these frameworks the determinate-experience problem simply vanishes.

\section{Solution I: Bohmian Mechanics}
\label{BM}

It is an established fact that according to BM observers do not play any role in measurement situations.\footnote{For significant literature about the role of observers in BM the reader may refer to \cite{Bohm:1952aa}, \cite{Durr:2013aa}, \cite{Durr:2009fk}, Chapter 9, \cite{Bricmont:2016aa}, Chapter 5, \cite{Passon2018}.} An experimenter in a Bohmian universe, i.e. a universe described by BM, can perform a measurement, leave the laboratory and be completely certain that, given the structure of the theory provided by its primitive ontology and dynamical laws, a definite outcome will be obtained. In this section I will schematically explain the basic elements which characterize the theory of measurement in the context of Bohmian mechanics.

BM provides a complete description of quantum systems via the introduction of additional variables, i.e. particles' positions, to the wave function. In this theory physical systems are represented by a couple $(Q(t), \psi(t))$, where $Q(t)=(Q_1(t),\dots, Q_N(t))$ describes an actual $N$-particle configuration at an arbitrary time $t$, and the latter is the configuration's wave function. As a consequence, Bohmian particles follow definite trajectories in physical space $\mathbb{R}^3$, contrary to the objects of standard quantum mechanics. 

The dynamics of BM is fully described by two equations of motion. The wave function $\psi$ evolves in $3N$-configuration space according to the unitary evolution provided by \eqref{SE}, whereas the particles' behaviour is governed by the so-called \emph{guidance} equation:
\begin{align}
\label{guide}
\frac{dQ_k}{dt}=\frac{\hbar}{m_k}\mathrm{Im}\frac{\psi^*\nabla_k\psi}{\psi^*\psi}(Q_1,\dots, Q_N)=v_k^{\psi}(Q_1,\dots, Q_N).
\end{align}
 
\noindent Evidently the vector velocity field on the r.h.s. depends on the wave function, whose role is to guide the motion of the particles in physical space on the one hand, and to determine the statistical distribution of the particles' positions on the other. Finally, the Born's rule is preserved in BM, making its predictions empirically equivalent to those of QM.\footnote{For the mathematical justification of this statement the reader should refer to \cite{Durr:2013aa}, Chapter 2, Secs. 4-7. In this regard, it is also worth noting that BM restores also \emph{a classical interpretation of quantum probabilities}. These are manifestation of our ignorance about the exact positions of the Bohmian particles and our inability to manipulate them. Thus, the maximum knowledge of particles' configurations at our disposal in BM is provided by $|\psi|^2$. For the specific treatment of the classical interpretation of probabilities in BM the reader should consult \cite{Bohm:1952aa}.}

To understand the Bohmian solution to the MP let us consider a wave function in a superposition of two states corresponding to the possible eigenstates of a two-valued operator $\mathcal{O}$, say  $L$ and $R$ with eigenvalues ``left'' and ``right'' (a situation analogous to our previous example of a particle's spin in the $z$-direction):
\begin{align*}
\psi=\frac{1}{\sqrt{2}}(\psi_L+\psi_R).
\end{align*}
According to this theory, a particular experimental situation is represented by a couple $(X,Y)\in\mathbb{R}^{3M}\times\mathbb{R}^{3(N-M)}$ where the former variable refers to the actual initial $M$-particle configuration of the subsystem under scrutiny, and the latter to the actual configuration of particle composing the environment, including the particle configuration of the experimental device which will register the effective measurement result.

Now, we assume that before the performance of the measurement at time $t=0$, the macroscopic pointer points in the neutral direction, i.e. the apparatus is in the ``ready'' state and the system has not yet interacted with it; 
thus, they are initially independent entities described by a product wave function. Running the experiment, the interaction between the system and the experimental device will be governed exclusively by the dynamical evolution provided by \eqref{SE} and \eqref{guide}. Since particles' positions of the measured system and the experimental apparatus are always well-defined, the measurement outcome will also be well-defined as a consequence of the theory's formalism. Hence, the configuration of particles $(X_0,Y_0)$ will deterministically evolve at a successive time $t>t_0$ into another configuration $(X_t,Y_t)$, corresponding to one of the possible eigenstate of the measured operator coupled with a definite state of the macroscopic experimental device, which will point in a definite direction expressed explicitly by $Y_t$. In measurement situations, as Passon notes, ``the wavefunction of the measurement apparatus will in
general be in a superposition state. The configuration however indicates the result of the measurement which is actually realized. That part of the wavefunction which ``guides'' the particle(s) can be reasonably termed the \emph{effective wavefunction}. All the remaining parts can be ignored, since they are irrelevant for the particle dynamics. As a result of decoherence effects [$\dots$], the probability that they will produce interference effects with the effective wavefunction is vanishingly small'' (\cite{Passon2018}, p. 191).

In our example, the probability to obtain $L$ is $|\alpha|^2$, similarly the probability to obtain $R$ is $|\beta|^2$; which result will be obtained depends exclusively on $(X_0, Y_0)$ and the dynamical laws \eqref{SE} and \eqref{guide}.\footnote{For technical details about the Bohmian theory of measurement and how BM solves the MP the reader may consult \cite{Durr:2009fk}, Chapter 9, \cite{Bohm:1952aa}, Part II, Section 2 and \cite{Maudlin:1995ab}.}
%

Taking into account the Bohmian theory of measurement, three remarks are in order:

\begin{enumerate}
   \item According to the Bohmian theory of measurement there are no superpositions of particles in physical space, therefore the physical situation described in \eqref{superpos} is avoided by construction. Consequently, macroscopic superposition are forbidden since the particle configuration representing an experimental situation always belongs to only one of the possible supports of the wave function. Thus, in this theory measurement results are functions of the primitive ontology and its dynamical evolution. The physical processes responsible for the macro-objectification of measurement outcomes are absolutely independent from external observers.
   \item In BM wave functions are not subjected to stochastic jumps: the wave function's collapse loses its fundamental role in the dynamics of the theory, being an \emph{effect} of the interaction between subsystem and environment. Once the experimenter looks at the final configuration of particles, then she finds the pointer in one of the possible (macroscopic) positions corresponding to one of the possible outcomes. Thus, within this theoretical framework by no means the observer induces or causes the result of a given measurement. 
   \item As already stressed at the outset of this section, in BM the experimenter can run the experiment and leave the lab: when she is absent the particle configuration will certainly evolve in a definite macroscopic state of the experimental device which is formally correlated to an eigenstate of the measured operator. Opening the lab's door, then, she updates her information about the outcome, but will not cause any modification to the physical situation. In fact, it is neither her act of observation, nor her conscious perception of the measurement outcome that is causally responsible for the observed result. Hence, if the MP is inherently related to the macro-objectification of macroscopic states from a microscopic quantum dynamics, then, it seems fair to claim that in BM the problem of the compatibility between the theory's dynamics and the  observer's definite experience of measurement results is simply dissolved. 
\end{enumerate}

Before concluding, it is useful to rectify some points present in Gao's treatment of the problem of measurement in BM. The author in \cite{Gao2018} claims that in this theory wave functions representing possible measurement outcomes do have tails spreading over all configuration space. This fact entails the possibility that the particle configuration constituting the macroscopic experimental device can be found in any point where the amplitude of the wave function is non-zero, and therefore, that this latter cannot be directly related with the value of the observed quantity. Again, this is not a serious threat for the Bohmian theory of measurement, and (above all) for the explanatory power of BM. The probability to find a particle in a given volume of space follows Born's rule, therefore, to find a Bohmian particle very far from the peak of its wave function is negligible, nearly null. Hence, the joint probability to find the \emph{entire} configuration of the particles composing the macroscopic apparatus in the tail of the pointer wave function is so extremely small to make this objection physically insignificant. Thus, it is fair to claim again that in BM macroscopic states not only are unambiguously generated by the microscopic dynamics, but also are well localized in physical space. Consequently, physical states representing measurement outcomes are those upon which mental states of observers supervene.

Finally, speaking about the form of psycho-physical supervenience for which observers' mental states supervene on the relative configuration of Bohmian particles, Gao claims that 
\begin{quote}
[i]f the mental state supervenes on the positions of the Bohmian particles, then an observer can in principle know the configuration of the Bohmian particles in her brain with a greater accuracy than that defined by the wave function. This will allow superluminal signaling and lead to a violation of the no-signaling theorem. (\cite{Gao:2017aa}, p. 8)
\end{quote}
The logic of the argument should be made clearer, since the consequent of the conditional follows from its antecedent only in certain peculiar cases. \cite{Durr:2013aa}, p. 74, in fact, underline that the cases in which an observer can know more information w.r.t the Born's rule, are those in which it is part of the system, as in the Wigner's friend scenario. However, this case is not explicitly taken into account in Gao's paper. Hence, from the supervenience of a certain observer's mental state -  which is generally not included in the system -, on the observation of a macroscopic pointer pointing in a definite direction, there are no reasons for which she should know with greater precision than $|\psi|^2$ the positions of the Bohmian particles. On the one hand, generally she is not part of the system and observes solely a macroscopic object, thus, she has no access whatsoever to the precise particles' positions; on the other, the mathematics of the theory simply forbids this possibility, as explicitly showed in \cite{Durr:2013aa} and \cite{Durr:2009fk}.  

\section{Solution II: the GRW process}
\label{GRW}

The GRW theory has been developed by the Italian physicists GianCarlo Ghirardi, Alberto Rimini and Tullio Weber in 1986 as a theoretical framework constructed to avoid the notorious defects of standard quantum mechanics. The fundamental assumption of this theory states that each microscopic, elementary massive component of a physical system can be subject to a random collapse, i.e. a spontaneous localization around its position in space with a certain mean frequency in time. It is worth stressing that these spontaneous processes are inherently stochastic: they are neither the result of physical interactions, nor of observers' consciousness. Nature, thus, is fundamentally indeterministic according to this theoretical framework. 

In order to succeed with this idea, the authors proposed to modify the fundamental dynamical law of the quantum mechanics \eqref{SE} adding stochastic terms referring to spontaneous random collapses of the wave function. Ghirardi, Rimini and Weber developed a mathematical model where they made explicit how the spontaneous localization works, i.e. explaining how a wave function is modified by this localization, where and when it occurs.\footnote{What follows is heavily influenced by \cite{Ghirardi:1997}, \cite{Ghirardi:2016}, \cite{Bassi:2003} and \cite{Allori:2008aa}. In this section I will stick to the bra-ket notation following the usual presentation of GRW theories.} 

More specifically, \cite{Bassi:2003} presented the basic assumptions of the GRW theory as follows:
\begin{enumerate}
   \item Each elementary component of any physical system composed by $N$ distinguishable massive particles is subject to random spontaneous localization processes with mean frequency rate $\lambda=10^{-16}$ sec$^{-1}$. 
   \item During the time interval between two spontaneous collapses, the system dynamically evolves according to the Schr\"odinger equation;
   \item The spontaneous localization process is described by:
   \begin{align*}
   |\psi\rangle\xrightarrow{evolution}\frac{|\psi_{\textbf{x}}^i\rangle}{\| |\psi_{\textbf{x}}^i\rangle\|}
   \end{align*}
   \noindent where $|\psi_{\textbf{x}}^i\rangle=L_{\textbf{x}}^i|\psi\rangle$. $L_{\textbf{x}}^i$ is a norm-reducing, positive, self-adjoint, linear operator in the $n$-particle Hilbert space $\mathcal{H}$, representing the localization of $i^{th}$ particle around the point $\textbf{x}$ (\cite{Bassi:2003}, p. 298). 
   
Let us consider the simplest case of a particle\footnote{In this version of GRW theory the term `particle' has not to be interpreted literally, since this theory is only about wave functions. We use a particle language only to remain stuck to the usual jargon. Bassi and Ghirardi made this point precise claiming that: ``when we speak of ``particles'' we are simply using the standard, somehow inappropriate, quantum mechanical language. Within dynamical reduction models particles are not point-like objects which move in space following appropriate trajectories according to the forces they are subjected to (as it is the case of, e.g., Bohmian mechanics). In dynamical reduction models, like in standard quantum mechanics, particles are represented just by the wave function which, in general, is spread all over the space. As we will see, the basic property of the models analyzed here is that, when a large number of ``particles'' interact with each other in appropriate ways [e.g. according to the GRW algorithm], they end up being always extremely well localized in space, leading in this way to a situation which is perfectly adequate for characterizing what we call a ``macroscopic object''. Thus, strictly speaking there are no particles in dynamical reduction models at the fundamental level; there is simply a microscopic, quantum, wave-like realm which gives rise to the usual classical realm at the macroscopic level'' (\cite{Bassi:2003}, p. 299).} moving on the $x$ axis, whose wave function $|\psi(\textbf{x})\rangle$ is not zero in an extended interval. Contrary to the wave function, the \emph{localization} function (which has a Gaussian form) $L_{\textbf{x}^*}$ will be different from $0$ only in a very narrow interval of amplitude $\delta$ whose center is $\textbf{x}^*$; within this interval $L_{\textbf{x}^*}$ has a constant value of $1/\sqrt{\delta}$. In this theory $\delta$ refers to the accuracy of the spontaneous localization and it is of order of $10^{-7}m$. Then, if the particle is randomly subjected to a spontaneous localization event in the neighbourhood of $\textbf{x}^*$, its wave function will be changed: $|\psi(\textbf{x})\rangle\longrightarrow N{L_{\textbf{x}^*}}|\psi(\textbf{x})\rangle$. The effect of the localization process is to make the initial wave function peaked within the interval $(\textbf{x}^*-\delta/2, \textbf{x}^*+\delta/2)$ and approximately null outside; here $N$ is a normalization factor, which ensures that the particle will be found in the interval in which has been localized if a position measurement were performed. 
As already pointed out above, GRW theory states also that the frequency $\lambda$ with which the collapses occur is $\lambda=10^{-16}s^{-1}$, meaning that a microscopic system is subjected to a stochastic localization ``on average, every hundred million years, while a macroscopic one undergoes a localization every $10^{-7}$ seconds'' (\cite{Ghirardi:2016}). 
   \item The probability density for the occurrence a of spontaneous localization at a point $\textbf{x}$ in space is $P_i($\textbf{x}$)=\| |\psi^i_{\textbf{x}}\rangle\|$. Consequently, the spontaneous collapses occur with greater probability where there is a high probability, according to standard quantum theory, to find a particle if a position measurement is performed. NB: this probability must not be confused with the usual quantum interpretation; rather, it is the probability for a certain spontaneous localization event to occur at a given point $\textbf{x}$ in space. 
\end{enumerate}

We may summarize the GRW algorithm in a nutshell: our initial wave function evolves according to (\ref{SE}) until a random localization event occurs, then, it will undergo a spontaneous collapse with center at $\textbf{x}^*$, which is chosen randomly with probability expressed as in point 4 above. After the random collapse, the wave function continues to follow SE until the next stochastic jump. 

After this sketchy (and naturally lacunose) presentation of the GRW process, it is time to explain how this theory solves the MP. Reconsider the superposition $|\psi\rangle=\frac{1}{\sqrt 2}(|\psi_L\rangle+|\psi_R\rangle)$ analyzed in the previous section, recalling that the states on the r.h.s. correspond to different orientations of a macroscopic pointer, which of course will be observed in a definite position after the performance of an experiment. Furthermore, since these pointers are composed by a huge number $N$ of quantum particles, we can write their wave functions as follows: $|\psi_L\rangle=|\psi_L(\textbf{x}_1, \textbf{x}_2, \dots, \textbf{x}_N)\rangle$ and similarly $|\psi_R\rangle=|\psi_R(\textbf{x}_1, \textbf{x}_2, \dots, \textbf{x}_N)\rangle$. $|\psi_L(\textbf{x}_1, \textbf{x}_2, \dots, \textbf{x}_N)\rangle$ will be different from zero only when the variables $(\textbf{x}_1, \textbf{x}_2, \dots, \textbf{x}_N)$ assume values in the neighborhood of $L$, analogously $|\psi_R(\textbf{x}_1, \textbf{x}_2, \dots, \textbf{x}_N)\rangle$ will be different from zero when $(\textbf{x}_1, \textbf{x}_2, \dots, \textbf{x}_N)$ are in the neighborhood of $R$.
NB: the distance between $L$ and $R$ is much greater than $\delta$, being the pointers macroscopically localized in disjoint regions. Then, the total wave function can be rewritten as follows:

\begin{align}
\label{grw_sup}
|\psi\rangle=\frac{1}{\sqrt 2}\Big{(} |\psi_L(\textbf{x}_1, \textbf{x}_2, \dots, \textbf{x}_N)\rangle+|\psi_R(\textbf{x}_1, \textbf{x}_2, \dots, \textbf{x}_N)\rangle\Big{)}.
\end{align}
\noindent Suppose now that a particle $\textbf{x}_k$, one the constituent of the macroscopic pointer, is subjected to a spontaneous localization. From the principles of the theory, we know that this random collapse can happen only in the spatial regions in which the particle $\textbf{x}_k$ has a value $\textbf{x}^*_L$ in the neighborhood of $L$ or $\textbf{x}^*_R$ in the neighborhood of $R$, and these two cases may happen with the very same probability. Suppose now that the event $\textbf{x}^*_L$ occurred, then multiplying $L_{\textbf{x}^*_L}(\textbf{x}_k)$ by \eqref{grw_sup} one obtains the suppression of the second summand $|\psi_R(\textbf{x}_1, \textbf{x}_2, \dots, \textbf{x}_N)\rangle$, since the localization function $L_{\textbf{x}^*_L}(\textbf{x}_k)$ is zero in the neighborhood of $R$. Thus, we obtained the localization and the macro-objectification of the pointer position: the localization of one of the pointer's constituents implies the spontaneous localization of all the particles composing it. We conclude, then, that the necessary and sufficient condition to obtain such macro-objectification amounts to require that one particle of its constituent will be subject to a stochastic localization event (see \cite{Ghirardi:1997}, p. 369). 

If a spontaneous localization event of one of the particles composing this system occurs, the GRW model imposes the condition according to which ``this particle is constrained to be either in the spatial region which it occupies when the state is $|\psi_L\rangle$, or in the one corresponding to $|\psi_R\rangle$. The linear superpositions is consequently transformed into a statistical mixture of state $|\psi_L\rangle$ and $|\psi_R\rangle$. Since the number of differently located particles is $N$, the reduction of states $|\psi_L\rangle$ and $|\psi_R\rangle$ occurs with a rate which is amplified by a factor $N$ with respect to the one, $\lambda_i$, which characterizes the elementary spontaneous localizations'' (\cite{Bassi:2003}, p. 305, notation adapted). It is worth noting a quantitative fact: the number $N$ of particles we are considering in these physical situations is of the order of Avogadro's number, so that for physical bodies with macroscopic size linear superpositions are suppressed extremely rapidly, avoiding in this manner the problem of measurement and implying that the theory is able to explain why macroscopic bodies have precise localizations in physical space. 

From this example one concludes that the MP is easily solved since the macro-objectification of classical devices is a mathematical consequence of the stochastic mechanism of the theory's dynamics. Thus, also in this case, conscious observers do not play any active role in the determination of measurement outcomes. Furthermore, the observer's mental state supervenes on a well-defined macroscopic state, which has been unambiguously obtained from clear physical processes. As in the case of Bohmian mechanics, in virtue of the clear dynamical structure characterizing the spontaneous localization process, this theory does not present an incompatibility between SE and the projection postulate, i.e GRW does not differentiate between measurement and non-measurement interaction, nor ill-defined notions as `measurement' or `observer' play a role in its physical content. 

However, although this theory solves effectively the MP, it is still solely about wave functions, thus, one may legitimate ask how macroscopic objects are obtained from them. This fact pushed Ghirardi, Grassi and Benatti to face this issue directly by proposing a variant of the theory adding a primitive ontology of matter density field, the so called GRWm theory. 
To describe the distribution of matter in physical space, GRWm introduces a continuous matter density field in addition to the wave function, representing it with the variable $m(\textbf{x},t)$, which is defined for every spatial point $\textbf{x}\in\mathbb{R}^3$ and every instant of time $t$.
Clearly, this field can have zero values in some point of space, i.e. where there is no matter.\footnote{For a accessible and precise presentation of the basic ideas of GRWm the reader may refer to \cite{Ghirardi:2016}, Section 11, for technical details see \cite{Ghirardi:1995aa}.} This theory, therefore, is about the dynamical behaviour of a matter density field in space and time, whose evolution is determined by the wave function. Another well-known variant of GRW theory implements a flash ontology (proposed in the first place in \cite{Bell:2004aa}, Chapter 22), where flashes are discrete material points in space-time. The wave function in this modification of GRW theory governs the evolution of flashes, which correspond bijectively to wave function collapses. In a nutshell this theory tells us that matter, i.e. the primitive ontology, is defined as the set of flashes in space at every instant of time; macroscopic bodies are therefore emergent from the distribution of such flashes. 

GRWm and GRWf solve the measurement problem exactly as GRW does, but spontaneous collapses now concern physical fields and flashes respectively, i.e matter in space, not only wave functions. Conscious observers, hence, play no role in the physics of these theories. More specifically, in these two variants of GRW theory macroscopic pointers, as any other macroscopic object, is composed by matter evolving in physical space according to precise dynamical laws, so that measurement outcomes, as in the case of BM, are just functions of their primitive ontologies and their dynamical evolution. The schema we have seen in the previous section replicates perfectly.

\vspace{2mm}

From what has been said, it follows that GRW theories solve the MP without any reference to external observers (and their consciousness). Exactly as in the case of BM, the experimenter can perform a measurement, leave the laboratory and be certain that the observation will produce a particular outcome. The missing information, hence, concerns which outcome has been produced. As in BM, the experimenter's observation has only the consequence to update her knowledge regarding the obtained result. 

As a corollary to this conclusion, one can state that in virtue of the primitive ontologies and the clear dynamical laws of GRW theories, the determinate-experience problem vanishes as in the Bohmian case.
 
\section{Conclusion}

In conclusion, the fundamental disagreement with Gao amounts to his claim that the MP reduces to a determinate-experience problem. According to the present author, he overlooks that the MP appears well before the introduction of observer's consciousness, being it a problem concerning the localization of macroscopic states from an obscure microscopic dynamics. More precisely, in this essay I have argued that the problem of measurement in the context of quantum mechanics is essentially a \emph{mathematical consequence} of the theory's formalism, translating into a physical problem consisting in the amplification on a macroscopic scale of microscopic superpositions. 

To summarize, in Section \ref{MP} I argued that it is not correct to define the MP a determinate-experience problem, and therefore, Gao's mentalistic reformulation is not helpful to understand the causes of the issue in the theory's formalism. In that section, it has been eventually shown that the physical definition of the MP does not \emph{reduce} to its mentalistic variant.

In Section \ref{BM} and \ref{GRW} I have considered the solutions to the MP offered respectively by Bohmian mechanics and GRW theories, clarifying some questionable statements contained in \cite{Gao:2017aa} about these theoretical frameworks. I have argued that according to these quantum theories measurement outcomes are just function of the theories' ontologies and dynamics. Moreover, it has been shown, explaining how measurement outcomes are obtained in these frameworks, that the act of observation by a conscious observer does not play any role in the macro-objectification of physical states. 

In other words, observers in a Bohmian universe or in a universe described by GRWm or GRWf, always have definite perceptions of macroscopic (well-localized) states in virtue of the primitive ontology and the dynamical laws of these theories, i.e. their mental states always supervene on physical state representing measurement outcomes. Thus, in these frameworks the determinate-experience problem simply vanishes.

Concluding one may also say that psychophysical supervenience is a substantial assumption in the philosophy of mind. Interactionist dualists, for instance, deny it. Nevertheless, in order to define the measurement problem and to understand how QM, BM and GRW theories solve it, there is absolutely no point to commit to a particular position in philosophy of mind, or to reintroduce again the mind into the physics of quantum theory.  
\vspace{5mm}

\textbf{Acknowledgements} 
I would like to thank Olga Sarno, Michael Esfeld and Davide Romano for helpful comments on this paper. I am grateful to the Swiss National Science Foundation for financial support (Grant No. 105212-175971).
\clearpage

\bibliographystyle{apalike}
\bibliography{PhDthesis}
\end{document}